\documentclass{article}
\usepackage{amsmath,amsthm}
\usepackage{amssymb,latexsym}
\usepackage[mathscr]{eucal}
\usepackage{setspace}
\usepackage{graphics}
\usepackage{array}
\usepackage{nccmath}
\usepackage{geometry}
\usepackage{float}
\usepackage{graphicx}
\usepackage{lineno}
\usepackage{lineno}
\newcommand{\tbf}[1]{\textbf{#1}}
\newcommand{\tit}[1]{\textit{#1}}

\begin{document}

\title{\LARGE \textbf{On the Role of Chapman's Hydrostatic Solar Wind Mechanism in Parker's Hydrodynamic Solar Wind Model}}

\author{{Bhimsen K. Shivamoggi} \\ 
University of Central Florida \\ 
Orlando, FL 32816-1364}

\date{}

\maketitle

\begin{center}
    \textbf{{Abstract}}
\end{center}

The role of Chapman's hydrostatic solar wind mechanism (resulting from a hydrostatic force balance condition) \cite{ref4} in Parker's hydrodynamic solar wind model \cite{ref6} is investigated by invoking the \textit{de Laval nozzle} analogy (Clauser \cite{ref14}, Parker \cite{ref15}) for the production of flow acceleration in the latter model.
The action of solar gravity in Parker's hydrodynamic solar wind model is shown to be geometrically equivalent to a \textit {renormalization} of the actual wind channel area and the renormalization factor is exactly Chapman's hydrostatic radial density profile \cite{ref4}, which is totally predicated on the hydrostatic force balance condition. This result appears to be traceable to the encapsulation of the solar gravity effects in Parker's hydrodynamic solar wind model by Chapman's hydrostatic solar wind mechanism \cite{ref4}, even beyond the coronal base. Furthermore, this result is shown to be robust by considering both isothermal gas and polytropic gas models as well as an n-dimensional ($n=$ 1, 2, 3) underlying space for the solar wind.

\newpage
\section{Introduction}

The solar wind is a hot tenuous plasma\footnote{The corona and the solar wind emerging from it are, in reality, fully ionized and highly electrically conducting and carry an associated remnant of solar magnetic field. The concomitant magnetization effects enable the solar wind to cause an enhanced outward transfer of angular momentum of the sun via the \tit{magnetic braking} process. Farther away from the sun, these magnetization effects also produce geomagnetic activity, like \tit{aurora borealis} in earth's magnetosphere.} outflowing continually from the sun, which carries off a huge amount of angular momentum from the sun while inflicting only a negligible loss of mass (Meyer-Vernet \cite{ref1}). The bulk of the solar wind is known to emerge from the coronal holes (Sakao et al. \cite{ref2}), and to fill the heliosphere (Dialynas et al. \cite{ref3}). Low to intermediate speed solar wind is believed to be caused by coronal heating along with high thermal conduction. Chapman \cite{ref4} argued that the corona is governed by a near hydrostatic force balance condition due to the strong binding of the corona by solar gravity, and hence Chapman's formulation has been termed Chapman's \tit{hydrostatic} model even though the radial density gradient implicit in the hydrostatic force balance nonetheless spawns a residual wind flow everywhere. Lamers and Casinelli \cite{ref5} indeed numerically demonstrated that the corona is almost in hydrostatic equilibrium not just at its base, but until close to the \textit{Parker sonic critical point} (\(r = r_*\)), where the wind flow speed equals the sound speed in the wind. 

However, away from the sun, as Parker \cite{ref6} pointed out, the thermal energy of the corona greatly exceeds the gravitational binding energy, so Chapman's \cite{ref4} hydrostatic force balance condition would become inaccurate, and the radial coronal flow needs to be \tit{fully} accounted for. Furthermore, Chapman's model gives finite values for the pressure at infinity in the interstellar space, which conflicts with zero pressure values needed there. Parker \cite{ref6} accordingly gave an ingenious \textit{hydrodynamic} model for the solar wind, which is predicated on the existence of a full-fledged solar wind outflowing continually from the sun. Furthermore, the solar wind is postulated to accelerate \textit{continuously} from subsonic speeds at the coronal base to supersonic speeds away from the sun \textit{via} conversion of the thermal energy in the wind beyond the coronal base into kinetic energy of the outward flow. Solar wind observations (Schrijver \cite{ref7}) indicated that the large-scale behavior of the solar wind, on the average, its local noisiness (Feldman et al. \cite{ref8}) notwithstanding, is apparently close to Parker's steady solar wind solution.\footnote{The Parker Solar Probe (Shivamoggi \cite{ref9}) has been providing 
significant information on the conditions in the inner solar corona (Fisk and Kasper \cite{ref10}, Bowen et al. \cite{ref11}, and others) some of which were at variance with previous belief (like the coupling of the solar wind with solar rotation (Kasper et al. \cite{ref12}) which was shown (Shivamoggi \cite{ref13}) to cause an enhanced angular momentum loss from the sun). The latest version of the Trans-Heliospheric Survey data (which is a compilation of \tit{in-situ} measurements from 13 different space probes gathered over a period of 60 years and over 3 orders of magnitude in distance from the sun (Brown et al. \cite{Brown})) showed a remarkable agreement of the observed proton speed in the inner heliosphere with Parker's steady solar wind model. This agreement was found also to largely hold even for an unsteady solar wind with fluctuations caused by wave motions and turbulence, \tit{albeit} long-time averaging the data (Maruca et al. \cite{Maruca}).}

On the other hand, the continuous acceleration of the solar wind to supersonic speeds, as described by Parker's hydrodynamic solar wind model \cite{ref6}, led to the surmise of a \textit{de Laval nozzle} type mechanism (Clauser \cite{ref14}, Parker \cite{ref15}) underlying the production of flow acceleration in the latter model. The effective de Laval nozzle associated with Parker's hydrodynamic solar wind model was also shown (Shivamoggi \cite{ref13}) to have a minimum cross-section area at the Parker sonic critical point (\(r = r_*\)), as expected.

Near the coronal base (\(r \ll r_*\)), Parker's steady solar wind solution \cite{ref6} reduces, as expected, to Chapman's hydrostatic solar wind solution \cite{ref4}. On the other hand, the numerical calculations of Lamers and Casinelli \cite{ref5} showed that the radial density profiles given by Chapman's hydrostatic force balance condition \cite{ref4} are almost identical to those given by Parker's hydrodynamic model \cite{ref6} (corresponding to the same temperature) in the \textit{whole} subcritical region (\(r < r_*\)). The purpose of this paper is therefore to investigate the role of Chapman's hydrostatic solar wind mechanism \cite{ref4} in Parker's hydrodynamic solar wind model \cite{ref6}. This is accomplished by invoking the de Laval nozzle analogy (Clauser \cite{ref14}, Parker \cite{ref15}) underlying the production of flow acceleration in Parker's hydrodynamic solar wind model \cite{ref6}. The robustness of this development is confirmed by considering both \tit{isothermal} gas and \tit{polytropic} gas models (Holzer and Axford \cite{Holzer16}, Parker \cite{ref16}, Shivamoggi and Pohl \cite{ref18}) as well as an n-dimensional ($n=$ 1, 2, 3) underlying space for the solar wind (Parker \cite{ref6}).\footnote{Consideration of the physics in an n-dimensional space in order to obtain insights into the corresponding physics in three-dimensional space is ubiquitous in theoretical physics. Such an approach goes back to Kaluza \cite{Kaluza}, who generalized Einstein's general relativity from four-dimensional space-time to five-dimensional space-time and determined the connections between electromagnetism and gravitation. Considerations of the turbulence dynamics in fluids in the hyperspace with $n>3$ (Kraichnan \cite{Kraichnan}, Shivamoggi \cite{Shivamoggi23}) provided similar insights into the corresponding dynamics in the three-dimensional space.}

\setcounter{section}{1} 
\section{Parker's Hydrodynamic Solar Wind Model}

Consider an ideal gas radial flow constituting the solar wind emanating from a central gravitating point mass representing the sun (Parker \cite{ref6}). The collisional mean free path in the solar corona is assumed to be small compared with the scale height of the corona, so the gas pressure may be taken to be nearly isotropic (Parker \cite{Parker21}). The flow is assumed to be steady and spherically symmetric so that the flow variables depend only on the distance \(r\) from the sun. Consider the gas flow to occur in a stream tube of cross-sectional area A\((r) = 4\pi r^2\) under isothermal conditions.

\vspace{1em}
The equations expressing the conservation of mass and momentum balance for this gas flow are (in usual notations),

\[
\frac{1}{\rho} \frac{d\rho}{dr} + \frac{1}{V_r} \frac{dV_r}{dr} + \frac{1}{A} \frac{dA}{dr} = 0 \tag{1}
\]

\[
\rho V_r \frac{dV_r}{dr} = -\frac{dp}{dr} - \rho \frac{GM_s}{r^2} \tag{2}
\]
where \(G\) is the gravitational constant, \(M_s\) is the mass of the sun, and \(p\) and \(\rho\) are related to each other \textit{via} the isothermal \tit{perfect gas} equation of state,

\[
p = \rho R T \equiv a^2 \rho \tag{3}
\]
\(a\) $ \equiv \sqrt{RT}$ is the constant speed of sound in the gas, and $R$ is the perfect gas constant. We assume that the flow variables, as well as their derivatives, vary continuously, so there are no shocks occurring anywhere in the region under consideration.

\vspace{1em}

Equations \((1) - (3)\) lead to:

\[
\frac{1}{V_r} \left( \frac{V_r^2}{a^2} - 1 \right) \frac{dV_r}{dr} = \left( \frac{1}{A} \frac{dA}{dr} - \frac{2r_*}{r^2} \right) \tag{4}
\]
where $r_* \equiv GM_s/2a^2$ locates the \textit{Parker sonic critical point}; here the wind flow speed equals the speed of sound in the wind.

\vspace{1em}
On noting A\((r) = 4\pi r^2\), equation \((4)\) becomes,

\[
\frac{1}{V_r} \left( \frac{V_r^2}{a^2} - 1 \right) \frac{dV_r}{dr} = \frac{2}{r^2} (r - r_*) \tag{5}
\]
which indicates the acceleration of subsonic wind flows to sonic speeds for \(r < r_*\), and the acceleration of wind flows to supersonic speeds for \(r > r_*\).

\section{The de Laval Nozzle Analogy}

The \textit{continuous} acceleration of the solar wind, as described by Parker’s hydrodynamic solar wind model \cite{ref10}, from subsonic speeds at the coronal base to supersonic speeds away from the sun led to the surmise of a \textit{de Laval nozzle} type mechanism (Clauser \cite{ref14}, Parker \cite{ref15}) underlying the production of flow acceleration in Parker’s hydrodynamic solar wind model \cite{ref6}.

It should be noted, however, that there are significant physical differences in the detailed manner by which the flow acceleration is produced in Parker's hydrodynamic solar wind model and the de Laval nozzle. In Parker's hydrodynamic solar wind model, thanks to the dominance of \tit{solar gravity} near the sun, the gas density drops drastically away from the sun, in the \tit{subsonic} $(v_r<a)$ region. This leads to the enhancement of the gas speed, the increase in the wind channel area outside the sun notwithstanding, to conserve the mass flux of the gas. So, in Parker's hydrodynamic solar wind model, the flow acceleration is \tit{compressibility} driven in the subsonic region. By contrast, in the de Laval nozzle, thanks to \tit{thermodynamics}, the gas density remains essentially constant in the \tit{subsonic} region, and the enhancement of the gas speed is primarily produced by the converging nozzle cross-section area, to conserve the mass flux of the gas. So, in the de Laval nozzle, the flow acceleration is \tit{flow-geometry} driven in the subsonic region. On the other hand, thanks again to \tit{thermodynamics} in the de Laval nozzle, the gas density drops drastically in the \tit{supersonic} region, so the flow acceleration (induced again by the conservation of mass flux) is \tit{compressibility} driven in this region.

\vspace{1em}

Indeed, if \(\mathcal{A} = \mathcal{A}(r)\) is the cross-sectional area of the \textit{effective de Laval nozzle} associated with Parker’s solar wind model, equation (4) leads to

\[
\frac{1}{V_r} \left( \frac{V_r^2}{a^2} - 1 \right) \frac{dV_r}{dr} = \left( \frac{1}{A} \frac{dA}{dr} - \frac{2r_*}{r^2} \right) \equiv \frac{1}{\mathcal{A}} \frac{d\mathcal{A}}{dr}. \tag{6}
\]

On using the boundary condition at the surface of the sun, given by \(r = r_0\),

\[
r = r_0 : \mathcal{A} = A \tag{7}
\]
equation (6) gives,

\[
\mathcal{A}(r) = A(r) e^{\frac{2r_*}{r_0} \left( \frac{r_0}{r} - 1 \right)}.\tag{8}
\]

(8) suggests a viable recipe to incorporate the solar gravity geometrically by renormalizing\footnote{The renormalization concept is standard practice in many-body physics. One example concerns \textit{Coulomb interactions} in a plasma. A test charge introduced in a plasma \textit{polarizes} it and acquires a \textit{shielding} cloud. It then becomes electrically invisible outside the cloud, and behaves like a neutral particle. So, the dielectric effects of a test charge may be transformed away \textit{via} the electrostatic \textit{shielding} process (Bellan \cite{ref20}), which constitutes a renormalization of Coulomb interactions in a plasma. On the other hand, in the context of turbulence dynamics in fluids, the accuracy of the perturbative expressions is improved by using a renormalization of the latter via a procedure called the \tit{direct interaction approximation} (DIA) (Kraichnan \cite{Kraichnan26}, see Shivamoggi and Tuovila \cite{Shivamoggi27} for mathematical insights into the DIA).} the actual wind channel cross-sectional area \textit{via} a multiplicative correction factor to yield the effective de Laval nozzle cross-sectional area \( \mathcal{A}(r) \). On the other hand, equation (6) may be viewed as an \textit{ansatz} to effectively excise solar gravity out of Parker’s hydrodynamic solar wind model \cite{ref10}.

\vspace{1em}

It may be noted that far away from the sun, (8) yields,

\[
\mathcal{A}(r) \approx A(r) e^{- \frac{2r_*}{r_0}} \tag{9}
\]
which indicates that the effective de Laval nozzle cross-sectional area \( \mathcal{A}(r) \) increases like the actual wind-channel area \( A(r) \) far away from the sun, where the solar gravity becomes negligible, as to be expected.

\vspace{1em}

Furthermore, putting \(A(r) = 4\pi r^2\), we have from (8),
\[
\mathcal{A}'(r) =  8\pi(r - r_*) e^{\frac{2r_*}{r_0} \left(\frac{r_0}{r} - 1\right)} \tag{10}
\]
which yields, 
\[
\mathcal{A}'(r) \lesseqgtr 0, \quad \text{if} \quad r \lesseqgtr r_*.
\tag{11}
\] 

In addition, noting from (10),
\[
\mathcal{A}''(r) = 8\pi \left[ 1 - \frac{2r_*}{r^2} (r - r_*) \right] e^{\frac{2r_*}{r_0} \left(\frac{r_0}{r} - 1\right)} \tag{12}
\]
we have,
\[
r \approx r_*: \quad \mathcal{A}(r) = \mathcal{A}(r_*) + \frac{1}{2} \mathcal{A}''(r_*) (\Delta r)^2 = e^{2(1 - r_*/r_0)} \left[ r_*^2 + 4\pi(\Delta r)^2 \right] \tag{13a}
\]
where,
\[
\mathcal{A}''(r_*) = 8 \pi e^{\frac{2 r_*}{r_o}\left(\frac{r_o}{r_*} -1 \right)} > 0, \hspace{0.1in} \Delta r \equiv r - r_*. \tag{13b}.
\]

(11) and (13) both confirm that the effective de Laval nozzle exhibits a minimum cross-sectional area at the Parker's sonic critical point, as to be expected.

Interesting physical implications of these results ensue by noting that the \textit{hydrostatic force balance} from equation (2), on using equation (3), gives (Chapman \cite{ref4}),
\[
-a^2 \frac{d\rho_h}{dr} - \frac{GM_s}{r^2} \rho_h = 0 \tag{14}
\]
where the subscript $h$ refers to hydrostatic conditions.

Equation (14), in conjunction with the boundary condition at the surface of the sun,
\[
r = r_0 : \quad \rho = \rho_0 \tag{15}
\]
yields
\[
\rho_h = \rho_0 e^{\frac{2r_*}{r_0} \left(\frac{r_0}{r} - 1\right)} \tag{16}
\]
where the subscript $0$ refers to the flow conditions at the surface of the sun.

Using (16), (8) may be rewritten as,
\[
\mathcal{A}(r) = A(r) \left[ \frac{\rho_h(r)}{\rho_0} \right]. \tag{17}
\]

(17) shows that the multiplicative correction factor needed to \textit{renormalize} the actual wind-channel area and incorporate the solar gravity geometrically, is precisely the \textit{Chapman hydrostatic density profile} \(\rho_h/\rho_0\). This indicates that the effects of solar gravity in Parker's hydrodynamic model \cite{ref6} are essentially encapsulated by Chapman's hydrostatic solar wind mechanism \cite{ref4} (even beyond the coronal base). This is corroborated by numerical calculations (Lamers and Cassinelli \cite{ref5}), which demonstrated that the corona is almost in hydrostatic equilibrium not just at its base, but until close to the Parker sonic critical point, so the radial density profiles associated with Chapman's hydrostatic model \cite{ref4} and Parker's hydrodynamic model \cite{ref6} (corresponding to the same temperature) are almost identical in the \textit{whole} subcritical region (\(r \leq r_*\)).

\section{Polytropic Gas Effects on the de Laval Nozzle Analogy}

\tit{In-situ} observations from solar and interplanetary probes (\tit{Helios 1} and \tit{Helios 2}) showed (Marsch et al. \cite{Marsch}) that solar wind expands \tit{non-isothermally} and does not also cool down as fast as that caused by an \tit{adiabatic} expansion (Boldyrev et al. \cite{ref19}). This may be traced to significant heating processes occurring both inside the corona and outside into the inner heliosphere, impairing adiabaticity in the wind. A more realistic model to describe this situation is to use the complete energy equation incorporating thermal conduction and coronal heating (Holzer et al. \cite{Holzer21}). However, thanks to the lack of knowledge about the energy flux carried by the waves from the photosphere (which are the basic source for the coronal heating), such an equation is inaccessible (Parker \cite{Parker22}). A practical approach then is to consider a \tit{polytropic gas} model (Holzer and Axford \cite{Holzer16}, Parker \cite{ref16}, Shivamoggi and Pohl \cite{ref18}), described by

\[
p = C \rho^\gamma \tag{18}
\]
where \(\gamma\) is the polytropic exponent, \(1 < \gamma < 5/3\), and \(C\) is an arbitrary constant. This \tit{ansatz} circumvents the inaccessible details involved with a complete energy equation concurrently stipulating nothing about the actual physical mechanisms underlying the coronal energy addition processes. The polytropic exponent \(\gamma\) characterizes the extent to which the solar coronal gas conditions deviate from adiabatic conditions (\(\gamma = 5/3\)) due to coronal heating \tit{via} thermal conduction and wave dissipation (Parker \cite{Parker23}) (see Appendix). Large-scale variations of the proton plasma density and temperature measured by the Parker Solar Probe within the inner heliosphere were found (Nicolaou et al. \cite{Nicolaou}) to follow a polytropic gas model with a polytropic exponent $1.5<\gamma<5/3$\footnote{The polytropic gas model with a single-value of $\gamma$ does not describe the whole coronal region accurately and $\gamma$ needs to be allowed to vary with distance from the sun to give better results (Parker \cite{Parker36}). Near the base of the corona, where the energy transfer is mainly due to thermal conduction, the best choice seems to be $\gamma =1$ (representing isothermal conditions). On the other hand, away from the coronal base, near-adiabatic conditions prevail, and the best choice seems to be $\gamma = 5/3$.} (see also Boxer et al. \cite{Boxer}), while short-scale fluctuations, probably due to turbulence in the solar wind, follow a polytropic gas model with a polytropic exponent $\gamma \sim 2.7$

\vspace{1em}

For a polytropic gas, the variations in the sound speed \(a\), given by,
\[
a^2 \equiv \frac{dp}{d\rho} \tag{19}
\]
on assuming the conservation of the total energy in the solar wind gas (Holzer \cite{ref17}),
\[
E \equiv \frac{v_r^2}{2} + \frac{a^2}{\gamma - 1} - \frac{GM_s}{r} = \text{const} \equiv \frac{a_0^2}{\gamma - 1}, \tag{20}
\]
are described by (Shivamoggi and Pohl \cite{ref18}),
\[
\frac{a^2}{a_0^2} = \frac{1 + 4 \alpha \frac{r_{*0}
}{r}}{1 + \alpha M^2} \tag{21}
\]
where $a_0$ is an effective sound speed (like the \tit{stagnation sound speed}\footnote{Stagnation thermodynamic values associated with a fluid particle result when the particle is brought to rest \tit{isentropically} and are produced by the energy exchanges occurring between the given particle and the surrounding fluid particles (Shivamoggi \cite{Shivamoggi}).} in a gas), which is just another way of expressing the total energy $E$. Here $M$ is the \tit{Mach number} of the flow,
\[
\quad M \equiv \frac{V_r}{a} \tag{22}
\]
and
\[
r_{*0} \equiv \frac{GM_s}{2a_0^2}, \quad \alpha \equiv \frac{\gamma - 1}{2}.
 \tag{23}
\]
\noindent The parameter $M$ has been introduced following Bondi's \cite{Bondi} conjecture in the context of the related spherically symmetric accretion flow model that it is the pertinent variable to formulate the polytropic gas case.

We note from equation (21),
\[
\frac{r_*}{r_{*_0}} = \frac{a_0^2}{a^2} = \frac{1+ \alpha M^2}{1+4 \alpha \frac{r_{*_0}}{r}}
\tag{24}
\]
which gives,
\[
r_* = r_{*_0}(1-3 \alpha)
\tag{25}
\]
(25) implies that $(1-3\alpha) > 0$ or $\gamma < 5/3$, which signifies the departure of the coronal gas from adiabatic conditions $(\gamma = 5/3)$ due to coronal heating activity.

On writing equation (4) as,
\[
\frac{1}{V_r} \left( \frac{V_r^2}{a^2} - 1 \right) \frac{dV_r}{dr} = \left[ \frac{1}{A} \frac{dA}{dr} - \frac{2}{r^2} \left( \frac{r_*}{r_{*0}} \right) r_{*0} \right], \tag{26}
\]
and using (24), equation (26) becomes
\[
\frac{1}{V_r} \left( \frac{V_r^2}{a^2} - 1 \right) \frac{dV_r}{dr} = 
\left[ \frac{1}{A} \frac{dA}{dr} - \frac{2r_{*0}}{r^2} 
\left\{ \frac{1 +  {\alpha} M^2}{1 + 4{\alpha}\frac{r_{*0}}{r}} \right\} \right]  \tag{27}
\]

If \(\mathcal{A} = \mathcal{A}(r)\) is the cross-sectional area of the effective de Laval nozzle associated with Parker's hydrodynamic polytropic solar wind model, we have
\[
\frac{1}{V_r} \left( \frac{V_r^2}{a^2} - 1 \right) \frac{dV_r}{dr} = \frac{1}{\mathcal{A}} \frac{d\mathcal{A}}{dr}. \tag{28}
\]
Equations (27) and (28) then lead to
\[
\frac{1}{\mathcal{A}} \frac{d\mathcal{A}}{dr} = \frac{1}{A} \frac{dA}{dr} - \frac{2r_{*0}}{r^2} 
\left\{ \frac{1 +{\alpha} M^2}{1 + 4{\alpha} \frac{r_{*0}}{r}} \right\} \tag{29}
\]

On using the boundary condition (7), equation (29) leads to,
\[
\mathcal{A}(r) = A(r) e^{-2r_{*0} \int_{r_0}^r \left\{ \frac{1 + {\alpha} M^2}{1 + 4 {\alpha}\frac{r_{*0}}{r}} \right\} \frac{1}{r^2} \, dr}. \tag{30}
\]
(30) suggests that the effective de Laval nozzle cross-sectional area \( \mathcal{A}(r) \) for the polytropic wind is again obtained from a \textit{renormalization} of the actual wind channel cross-sectional area \( A(r) \) \textit{via} a multiplicative correction factor incorporating the solar gravity in the Parker hydrodynamic polytropic solar wind model.

Furthermore, on noting \( A(r) = 4\pi r^2 \), and using (24), we obtain from (30)
\[
\mathcal{A}'(r) = 8\pi \left[ r - r_{*0} \left( \frac{1 + \alpha M^2}{1 + 4\alpha \frac{r_{*0}}{r}} \right) \right]
e^{-2r_{*0} \int_{r_0}^r \left( \frac{1 + \alpha M^2}{1 + 4\alpha \frac{r_{*0}}{r}} \right) \frac{1}{r^2} \, dr} \tag{31}
\]
(31) implies,
\[
\mathcal{A}'(r) \lesseqgtr 0, \quad \text{if} \quad r \lesseqgtr r_{*} \tag{32}.
\]
In addition, on noting from (31),
\[
\mathcal{A}''(r) = 8\pi \left[ 1 - r_{*0} \left( \frac{1 + \alpha M^2}{1 + 4\alpha \frac{r_{*0}}{r}} \right)' \right]
e^{-2r_{*0} \int_{r_0}^r \left( \frac{1 + \alpha M^2}{1 + 4\alpha \frac{r_{*0}}{r}} \right) \frac{1}{r^2} \, dr}
\]
\[
+ 8\pi \left[ r - r_{*0} \left( \frac{1 + \alpha M^2}{1 + 4\alpha \frac{r_{*0}}{r}} \right) \right]
 \left[ -\frac{2r_{*0}}{r^2} \left( \frac{1 + \alpha M^2}{1 + 4\alpha \frac{r_{*0}}{r}} \right) \right]
 e^{-2r_{*0} \int_{r_0}^r \left( \frac{1 + \alpha M^2}{1 + 4\alpha \frac{r_{*0}}{r}} \right) \frac{1}{r^2} \, dr} \tag{33}
\]
and using the result (Shivamoggi and Pohl \cite{ref18}),

\[
r = r_{*} = r_{*0} \left( \frac{1+\alpha M^2}{1+4 \alpha \frac{r_{*0}}{r}} \right) : \quad M^2 = 1, \quad \frac{dM^2}{dr} > 0 \quad \text{or} \quad r_{*0} \left( \frac{1+\alpha M^2}{1+4\alpha \frac{r_{*0}}{r}} \right)' < 1 \tag{34}
\]
we obtain,

\[
\mathcal{A}''(r_*) > 0 \tag{35}
\]

Thus,
\[
r \approx r_* : \mathcal{A}(r) \approx \mathcal{A}(r_*) + \frac{1}{2} \mathcal{A}''(r_*) (\Delta r)^2 \tag{36}
\]
which, on using (35), confirms that the effective de Laval nozzle, for a polytropic wind, exhibits again a minimum cross-sectional area at the Parker sonic critical point, as anticipated.

In order to further consider the physical implications of the above results, note that the hydrostatic force balance, from equation (2), gives:
\[
-\frac{dp_h}{dr} - \frac{GM_s}{r^2} \rho_h = 0. \tag{37}
\]
Equation (37), on using (19), leads to,
\[
\frac{1}{\rho_h} \frac{d\rho_h}{dr} = -\frac{1}{r^2} \left( \frac{GM_s}{a^2} \right). \tag{38a}
\]
Rewriting equation (38a) as,
\[
\frac{1}{\rho_h} \frac{d\rho_h}{dr} = -\frac{2}{r^2} \left( \frac{GM_s}{2a_0^2} \right) \left( \frac{a_0^2}{a^2} \right), \tag{38b}
\]
and using (21)-(23), we obtain,
\[
\frac{1}{\rho_h} \frac{d\rho_h}{dr} = -\frac{2}{r^2} r_{*0} 
\left\{ 
\frac{1 + \alpha M^2}{1 + 4 \alpha \frac{r_{*0}}{r}}
\right\}. \tag{39}
\]

Equation (39), in conjunction with the boundary condition (15) at the surface of the sun, yields,
\[
\rho_h = \rho_0 e^{-2r_{*0} \int_{r_0}^r 
\left\{ 
\frac{1 + \alpha M^2}{1 + 4 \alpha \frac{r_{*0}}{r}}
\right\} \frac{1}{r^2} \, dr}. \tag{40}
\]

Using (40), (30) may be rewritten as,
\[
\mathcal{A}(r) = A(r) \left[ \frac{\rho_h(r)}{\rho_0} \right] \tag{41}
\]
which is identical to the result (17), deduced before for the isothermal gas.

(41) implies that the multiplicative correction factor needed to \textit{renormalize} the actual polytropic-wind channel area and incorporate the solar gravity geometrically, is precisely the \textit {polytropic Chapman hydrostatic} radial density profile \( \rho_h / \rho_0 \), as in the isothermal gas case. This indicates that the effects of solar gravity in Parker's hydrodynamic solar wind model \cite{ref6} are essentially encapsulated by Chapman's hydrostatic solar wind mechanism \cite{ref4} (even beyond the coronal base), also in the polytropic case, hence demonstrating the robustness of this result.

\section{Parker's Hydrodynamic Solar Wind Model and the\\ de Laval Nozzle Analogy in an n-dimensional Space}

Parker \cite{ref6} pointed out that the flow acceleration process in Parker's hydrodynamic solar wind model depends on the dimension of the underlying space in a qualitative manner. A case in point is the one-dimensional space, for which the wind flow velocity was found (Parker \cite{ref6}) to remain subsonic and reach the sound speed asymptotically at an infinite distance from the sun\footnote{The one-dimensional "\tit{purely subsonic}" solar wind flow may mimic flows in some isolated magnetic flux tubes embedded in a magnetic field-free fluid. These flux tubes are manifestations of extreme inhomogeneity of the magnetic field near the solar surface, and possibly in the entire convection zone (Spruit and Roberts \cite{Spruit}), and act like conduits, trapping plasma inside them.}. It is therefore pertinent to investigate the impact of space dimension on the de Laval nozzle analogy as well, and hence on the functionality of Chapman's hydrostatic solar wind mechanism in Parker's hydrodynamic solar wind model in an n-dimensional space.

In an n-dimensional space, ($n =$ 1, 2, 3), equation (4) becomes
\[
\frac{1}{V_r} \left( \frac{V_r^2}{a^2} - 1 \right) \frac{dV_r}{dr} = \frac{1}{A_n}\frac{dA_n}{dr} - \frac{2 r_*}{r^2} 
\tag{42}
\]
where $A_n(r)$ is the area of the n-dimensional "sphere",
\[
A_n(r) = \beta_n r^{n-1}, \hspace{.1in} \beta_n \equiv \frac{2 \pi^{n/2}}{\Gamma(n/2)}.
\tag{43}
\]
(43) yields,
\[
\left.
\begin{aligned}
&n = 1: A_1 = \beta_1, \hspace{.1in} \beta_1 = 2 \\
&n=2: A_2 = \beta_2 r, \hspace{.1in} \beta_2 = 2 \pi \\ 
&n=3: A_3 = \beta_3 r^2, \hspace{.1in} \beta_3 = 4\pi
\end{aligned} \right\}
\tag{44}
\]

The Parker sonic critical point $r=r_{*_n}$, from equation (42), is given by

\[
r_{*_n} \equiv \frac{2 r_*}{n-1}.
\tag{45}
\]
which yields,
\[
\left.
\begin{aligned}
    &n=1: r_{*_1} \Rightarrow \infty\\
    &n=2: r_{*_2} = 2 r_*\\
    &n=3: r_{*_3} = r_*
\end{aligned}
\right\}
\tag{46}
\]
so the Parker sonic critical point moves farther away from the sun, as the dimension of the underlying space decreases, and it goes to infinity for a one-dimensional space (see Figure 1 below) signifying the \tit{asymptotic} attainment of the sonic speed by the solar wind for the latter case. This appears, as indicated by equation (42), to be due to the enhanced retardation of the solar wind caused by the augmented solar gravity effects, as the underlying space dimension decreases.

If $\mathcal{A}_n = \mathcal{A}_n(r)$ is the cross-sectional area of the effective de Laval nozzle associated with Parker's hydrodynamic solar wind model in an n-dimensional space, we have
\[
\frac{1}{V_r} \left( \frac{V_r^2}{a^2} - 1 \right) \frac{dV_r}{dr} = \frac{1}{\mathcal{A}_n} \frac{d \mathcal{A}_n}{dr}.
\tag{47}
\]

Using the boundary condition,
\[
r = r_o: \hspace{.1in} \mathcal{A}_n = A_n
\tag{48}
\]
equations (42) and (47) lead to
\[
\mathcal{A}_n(r) = A_n(r) e^{\frac{2 r_*}{r_o}\left(\frac{r_0}{r} -1 \right)}
\tag{49}
\]
The effective de Laval nozzle cross-section area profiles given by (49) and (44) are sketched in Figure 1, where
\[
\hat{\mathcal{A}}_n(\hat{r}) \equiv \left( \frac{e^{\frac{2 r_*}{r_o}}}{\beta_n r_*^{n-1}} \right) \mathcal{A}_n(r) = \hat{r}^{n-1}e^{2/\hat{r}}, \hspace{.1in} \hat{r} \equiv \frac{r}{r_*}.
\]
\begin{figure}[H]
    \begin{center}
    \includegraphics[width=0.6\textwidth]{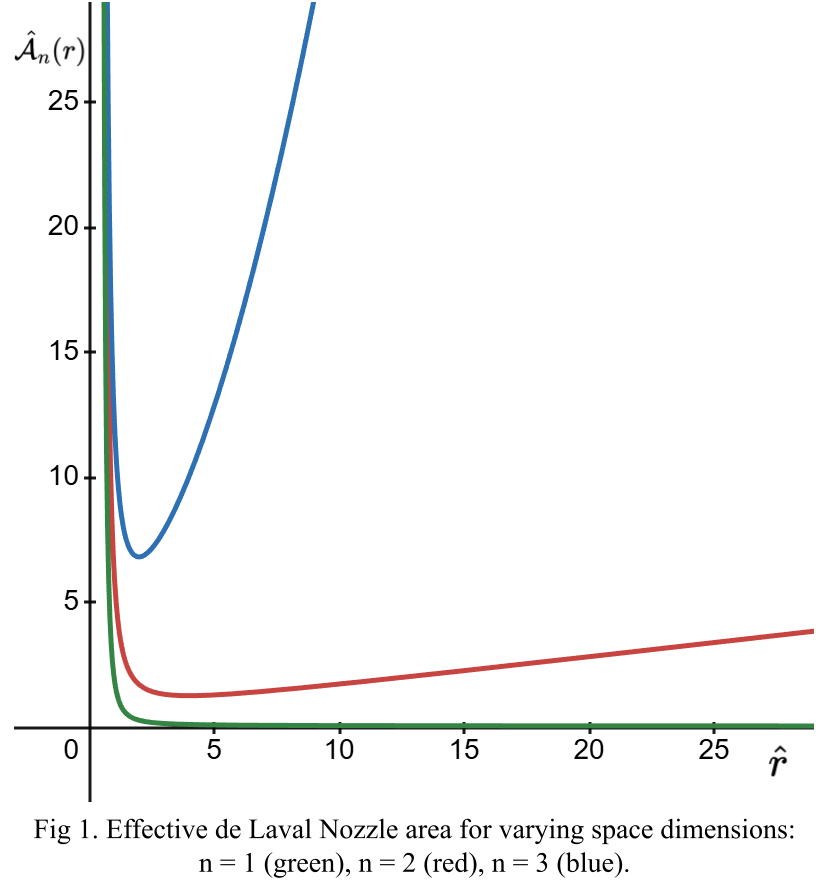}
    \end{center}
\end{figure}
Observe that the effective de Laval nozzle is purely \tit{converging} in a one-dimensional space, which is consistent with,
\begin{itemize}
    \item the recession of the Parker sonic critical point to infinity, as indicated by (46), for $n=1$;
    \item Parker's result \cite{ref6} showing a purely subsonic flow regime for the solar wind, for $n=1$.
\end{itemize}

On the other hand, the hydrostatic force balance, underlying Chapman's model, yields for all $n$, (see (16)),
\[
\rho_h = \rho_o e^{\frac{2 r_*}{r_o}\left(\frac{r_0}{r} -1 \right)}
\tag{50}
\]

Using (50), (49) becomes
\[
\mathcal{A}_n(r) = A_n(r) \left[ \frac{\rho_h(r)}{\rho_o} \right] \tag{51}
\]
which reduces to (17), for $n=3$. (51) shows that the multiplicative correction factor needed to renormalize the actual wind channel area and incorporate the solar gravity geometrically, is precisely the Chapman \cite{ref4} hydrostatic radial density profile $\rho_h/\rho_o$. This indicates that the effects of solar gravity in Parker's hydrodynamic solar wind model \cite{ref6} are essentially encapsulated by Chapman's hydrostatic solar wind mechanism \cite{ref4} (even beyond the coronal base), also in an n-dimensional space ($n=$ 1, 2, 3).

\section{Discussion}

Thanks to the strong binding of the corona by the solar gravity, Chapman \cite{ref4} argued that the corona is governed by a near hydrostatic force balance condition, and hence gave an essentially \tit{hydrostatic} model for the inner corona. Parker \cite{ref6} pointed out that Chapman's hydrostatic model \cite{ref4} becomes inaccurate away from the sun because the radial coronal flow needs to be \tit{fully} accounted for, and gave a full-fledged \textit{hydrodynamic} model for the solar wind, to supersede Chapman's hydrostatic model \cite{ref4}, and reduce to the latter near the coronal base (\(r \ll r_*\)), as expected. On the other hand, the numerical calculations of Lamers and Casinelli \cite{ref5} showed that the radial density profiles given by Chapman's hydrostatic model \cite{ref4} are almost identical to those given by Parker's hydrodynamic model \cite{ref6} (corresponding to the same temperature) in the \textit{whole} subcritical region (\(r \lesssim r_*\)). This warrants an investigation of the deeper role of Chapman's hydrostatic solar wind mechanism \cite{ref4} in Parker's hydrodynamic solar wind model \cite{ref6}. In this paper, we have sought to accomplish this by invoking the \textit{de Laval nozzle analogy} (Clauser \cite{ref14}, Parker \cite{ref15}) underlying the production of flow acceleration in Parker's hydrodynamic solar wind model \cite{ref6}, and have shown that the action of solar gravity in Parker's hydrodynamic solar wind model \cite{ref6} is geometrically equivalent to \tit{renormalization} of the actual wind channel area and the renormalization factor is precisely Chapman's hydrostatic radial density profile. So, the solar gravity effects in Parker's hydrodynamic solar wind model \cite{ref6} are essentially encapsulated by Chapman's hydrostatic model \cite{ref4}, even beyond the coronal base. The robustness of these results is confirmed by considering both isothermal gas and polytropic gas models (Holzer and Axford \cite{Holzer16}, Parker \cite{ref16}, Shivamoggi and Pohl \cite{ref18}) as well as an n-dimensional ($n=$ 1, 2, 3) underlying space for the solar wind.

\section{Acknowledgements}

This work got started during my sabbatical leave at California Institute of Technology. My thanks are due to Professor Shrinivas Kulkarni for his enormous hospitality and helpful discussions. I am thankful to Professor Earl Dowell for his valuable remarks and suggestions, and to Alexander Manganais for his help with Figure 1. I am thankful to the referees for their valuable remarks.

\section{Appendix}
For a polytropic process, the ratio of the energy transferred into the given system as heat and the energy transferred out of the system as work remains constant (Chandrasekhar \cite{Chandrasekhar}):

\begin{equation}
c \equiv \frac{dQ}{dT} = \hspace{.1in} \textnormal{const.}
\tag{A.1}
\end{equation}
which implies that 
\begin{itemize}
    \item $c=0$, for an adiabatic process,
    \item $c \Rightarrow \infty$, for an isothermal process.
\end{itemize}
Using the relation (in usual notation),

\begin{equation}
    dQ = dU - p \frac{d\rho}{\rho^2} = C_v dT - p \frac{d \rho}{\rho^2}
    \tag{A.2}
\end{equation}
$U$ being the internal energy, and the perfect gas equation of state,

\begin{equation}
    p = \rho R T
    \tag{A.3}
\end{equation}
(A.1) leads to

\begin{equation}
\frac{dp}{p} - \gamma \frac{d \rho}{\rho} = 0
\tag{A.4}
\end{equation}
or

\begin{equation}
    \frac{p}{\rho^\gamma} = \hspace{.1in} \textnormal{const.}
    \tag{A.5}
\end{equation}
where $\gamma$ is the polytropic exponent,

\begin{equation}
    \gamma \equiv \frac{C_p - c}{C_V -c}.
    \tag{A.6}
\end{equation}
(A.6) implies,
\begin{itemize}
    \item $\gamma = \frac{C_p}{C_V}$, for an adiabatic process.
    \item $\gamma = 1$, for an isothermal process.
\end{itemize}


\end{document}